%% file: main.tex
\documentclass[10pt,twocolumn,letterpaper]{article}

\usepackage{iccv}
\usepackage{times}
\usepackage{epsfig}
\usepackage{graphicx}
\usepackage{amsmath}
\usepackage{amssymb}
\usepackage{xcolor}

\usepackage[breaklinks=true,bookmarks=false]{hyperref}

\usepackage{color}

\iccvfinalcopy 

\def\iccvPaperID{****} 

\ificcvfinal\fi

\input{defs}
\usepackage[accsupp]{axessibility}

\begin{document}

\title{\ourframework: Automatic ReLU Replacement for Fast Private Network Inference}

\author{
$^{+}$Hongwu Peng$^{1}$
\and $^{+}$Shaoyi Huang$^{1}$
\and $^{+}$Tong Zhou$^{2}$ 
\and Yukui Luo$^{2}$ 
\and Chenghong Wang$^{4}$ 
\and Zigeng Wang$^{1}$$\thanks{Z. Wang is now affiliated with Walmart Global Tech, Sunnyvale, CA. }$
\and Jiahui Zhao$^{1}$ 
\and Xi Xie$^{1}$ 
\and Ang Li$^{5}$ 
\and Tony Geng$^{6}$ 
\and Kaleel Mahmood$^{1}$
\and Wujie Wen$^{3}$ 
~~  Xiaolin Xu$^{2}$ 
~~ Caiwen Ding$^{1}$  \\
{\tt\small $^{+}$These authors contributed equally. } \\
$^{1}$University of Connecticut \quad 
$^{2}$Northeastern University \quad
$^{3}$ North Carolina State University  \quad \\
$^{4}$ Duke University \quad
$^{5}$ Pacific Northwest National Laboratory \quad
$^{6}$ University of Rochester ~~~\\
{\tt\small \{hongwu.peng, shaoyi.huang, zigeng.wang, jiahui.zhao, xi.xie, 
 kaleel.mahmood\}@uconn.edu} \\
{\tt\small \{zhou.tong1, luo.yuk, x.xu\}@northeastern.edu, cw374@duke.edu,
ang.li@pnnl.gov
} \\
{\tt\small  tgeng@ur.rochester.edu, wwen2@ncsu.edu,
 caiwen.ding@uconn.edu
}
}

\maketitle


\input{Sections/Abstract}
\input{Sections/Sec1_Introduction}

\input{Sections/Sec2_Background}
\input{Sections/Sec3_AutoReP}

\input{Sections/Sec4_experiment}

\input{Sections/Sec5_conclusion}



{\small
\bibliographystyle{ieee_fullname}
\bibliography{main}
}



\end{document}

%% file: defs.tex
\newcommand{\ourframework}{AutoReP\xspace}

\usepackage{ amssymb }
\usepackage{multicol}
\usepackage{wrapfig}
\usepackage{subfig}

\usepackage{xspace}
\usepackage{xcolor}
\usepackage{amsmath, amsfonts, amsthm} 
\usepackage{graphicx}
\usepackage{float} 
\usepackage{hyperref}
\usepackage{wrapfig}
\usepackage{mathrsfs}
\usepackage{multicol}
\usepackage{balance}
\usepackage{mathtools}
\usepackage{tikz}

\usepackage{tabularx} 
\usepackage{adjustbox} 
\usepackage{multirow} 

\usepackage{complexity}
\usepackage{relsize}
\usepackage{graphicx}
\usepackage{epsfig}
\usepackage{flushend}
\usepackage{epstopdf}
\usepackage{longtable}
\usepackage{stmaryrd}
\usepackage{pstricks}
\usepackage{pst-tree}
\usepackage{epstopdf}
\usepackage{lipsum}
\usepackage{color}
\usepackage{framed}
\usepackage[normalem]{ulem}
\usepackage{float}
\usepackage{caption}
\usepackage{fancyvrb}
\usepackage{bm}

\usepackage{algorithmic}
\usepackage{algorithm}

\newcommand{\eat}[1]{}

\newcommand{\ignore}[1]{}

\newcommand{\share}[1]{\llbracket #1 \rrbracket}
\newcommand{\bshare}[1]{\langle #1 \rangle}

\newcommand{\pprotocol}[5]{
{
\begin{center}
 \advance\leftskip-2cm
 \advance\rightskip-2cm
\setlength{\protowidth}{\textwidth}
\addtolength{\protowidth}{\intextsep}
\fbox{
        \hbox{\quad
        \begin{minipage*}{\protowidth}
        #5
        \end{minipage*}
        \quad}
        }
        \caption{\label{#3} #2}
\end{center}

} }

\usepackage{tikz}
\usetikzlibrary{tikzmark}

%% file: Sections/Abstract.tex
\begin{abstract}

The growth of the Machine-Learning-As-A-Service (MLaaS) market has highlighted clients' data privacy and security issues. Private inference (PI) techniques using cryptographic primitives offer a solution but often have high computation and communication costs, particularly with non-linear operators like ReLU. Many attempts to reduce ReLU operations exist, but they may need heuristic threshold selection or cause substantial accuracy loss. This work introduces \ourframework, a gradient-based approach to lessen non-linear operators and alleviate these issues. It automates the selection of ReLU and polynomial functions to speed up PI applications and introduces distribution-aware polynomial approximation (DaPa) to maintain model expressivity while accurately approximating ReLUs.
Our experimental results demonstrate significant accuracy improvements of 6.12\% (94.31\%, 12.9K ReLU budget, CIFAR-10), 8.39\% (74.92\%, 12.9K ReLU budget, CIFAR-100), and 9.45\% (63.69\%, 55K ReLU budget, Tiny-ImageNet) over current state-of-the-art methods, e.g., SNL. Morever, \ourframework is applied to EfficientNet-B2 on ImageNet dataset, and achieved 75.55\% accuracy with 176.1 $\times$ ReLU budget reduction. The codes are shared on Github\footnote{\url{https://github.com/HarveyP123/AutoReP}}. 
\end{abstract}

%% file: Sections/Sec1_Introduction.tex

\section{Introduction}




The MLaaS market has seen significant growth in recent years, with many MLaaS platform providers eatablished, e.g, AWS Sagmaker~\cite{amazonsagemaker}, Google AI Platform~\cite{bisong2019overview}, Azure ML~\cite{team2016azureml}.
However, most MLaaS solutions require clients to share their private input, compromising data privacy and security. 
Private inference (PI) techniques, have emerged to preserve data and model confidentiality, providing strong security guarantees.
The existing highly-secure PI solutions usually use cryptographic primitives include multi-party computation (MPC)~\cite{goldreich1998secure,bonawitz2017practical,gilad2016cryptonets} and homomorphic encryption (HE)~\cite{juvekar2018gazelle,brutzkus2019low,dathathri2019chet,kim2022secure}. Recently, MPC-based PI becomes popular as it supports large-scale networks by partitioning
the inference between clients and MLaaS providers.

\begin{figure}[t]
    \centering
      \includegraphics[width=0.96\linewidth]{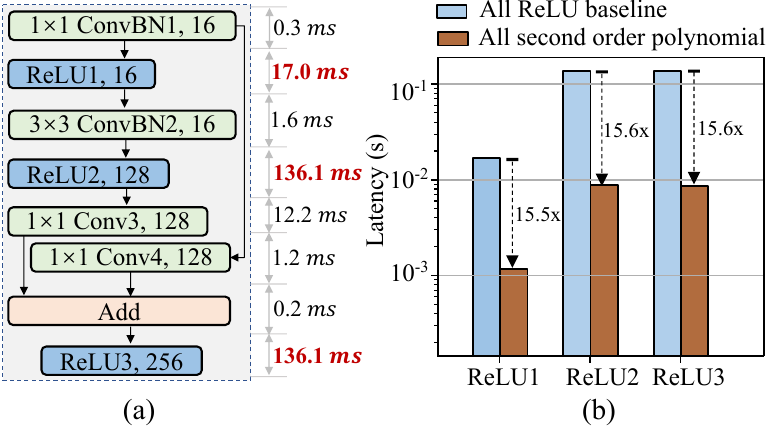}
  \captionof{figure}{Network bandwidth: 1 GB/s. (a) Latency breakdown of Wide-ResNet 22-8 operators under 2PC PI setup. (b) Latency reduction with polynomial replacement.  
  }
    \label{fig:motivation}
\end{figure} 

The main challenge of applying cryptographic primitives in
PI comes from the non-linear operators (e.g., ReLU), which introduces ultra-high computation and communication overhead. Fig.~\ref{fig:motivation} (a) shows that ReLU dominates the PI latency, i.e., up to 18.6 $\times$ than the combination of convolutional (Conv) and batch normalization (BN) operations. Reducing these ReLU operators could bring latency reduction, as highlighted in Fig.~\ref{fig:motivation} (b).
Centered by this observation, several approaches have been discussed,
including replacing ReLUs with linear functions (e.g., SNL~\cite{cho2022selective}, DeepReduce~\cite{jha2021deepreduce}) or low degree polynomials (e.g., Delphi~\cite{mishra2020delphi}, SAFENet~\cite{lou2020safenet}), designing neural architectures with fewer ReLUs 
(e.g., CryptoNAS~\cite{ghodsi2020cryptonas} and Sphynx~\cite{cho2022sphynx}), and ultra-low bit representations
(TAPAS~\cite{sanyal2018tapas}, XONN~\cite{riazi2019xonn}). 
However, these techniques (i) 
require a heuristic threshold selection on ReLU counts, therefore can not effectively perform design space exploration on ReLU reduction, resulting in sub-optimal solutions, 
and (ii) result in a significant accuracy drop on large networks and datasets such as ImageNet, hence are not scalable as the number of ReLUs or the number of bits decreases.

We argue that the root cause of the limitations is the disjointedness of non-linear operator reduction and model expressivity in this emerging field. We aim to systematically solve the efficient PI problem by answering two gradually advancing questions:  \textcircled{1} \textit{Which} non-linear operators should be replaced, and  \textcircled{2} \textit{What} to be replaced with 
to maintain a high model accuracy and expressivity, especially for large DNNs and datasets? 
In this work, 
we introduce a gradient-based \textbf{automatic ReLU replacement (\ourframework)} framework that incorporates joint \textit{fine-grained replacement 
policy (addressing  \textcircled{1})
} and \textit{polynomial approximation (addressing  \textcircled{2})}. Our framework could simultaneously reduce the non-linear operators  and maintain high model accuracy and expressivity. 
In summary, our contributions are as follows:
\begin{enumerate}
\item  We 
introduce a \textbf{parameterized discrete indicator function},
co-trained with model weights until convergence.
Our approach allows for fine-grained selection of ReLU and polynomial functions at the pixel level, resulting in a more optimized and efficient model.
\item We present a \textbf{hysteresis loop} update function to enhance the stability of the binarized ReLU replacement training process, which enables a recoverable and stable replacement and leads to better convergence and higher accuracy. 
\item  
Our proposed method, \textbf{distribution-aware polynomial approximation (DaPa)}, offers a novel solution to the problem of accurately approximating ReLUs using polynomial functions under specific feature distributions. By minimizing the structural difference between the original and replaced networks and maintaining high model expressivity.
\end{enumerate}

Experimental results show that our \ourframework (ResNet-18) achieves 74.92\% accuracy with 12.9K ReLU budget, 8.39\% higher than SNL~\cite{cho2022selective}, with 1.7x latency reduction, on CIFAR-100. For 73.79\% accuracy, \ourframework requires only 6K ReLUs, an 8.2x reduction in ReLU budget vs. SNL~\cite{cho2022selective}. 
When applied to the larger EfficientNet-B2 on the ImageNet dataset, \ourframework achieved an accuracy of 75.55\% with a significant reduction of 176.1 $\times$ in ReLU budget.

%% file: Sections/Sec2_Background.tex

\section{Background and Related Work}
\label{sec:background}

\subsection{Threat Model and Cryptographic Primitives}

\noindent\textbf{2PC setup.} 
In this paper, we explore a two-party secure computing (2PC) protocol for MLaaS, leveraging prior work~\cite{demmler2015aby, kamara2011outsourcing}. The protocol lets the client outsource confidential inputs to two servers, who use a 2PC protocol to compute a function securely without revealing intermediate information or results. This approach can scale to enable secure computation for multiple clients with confidential inputs, as demonstrated in~\cite{demmler2015aby}.

\noindent\textbf{Threat model.} 
Here, we focus on an admissible adversary~\cite{mohassel2017secureml} who can compromise one server at a time, which aligns with the non-colluding server assumption in MPC. Our security model assumes semi-honest behavior~\cite{patra2021aby2,choi2019hybrid,husted2013gpu,zhang2013picco}, where the adversary follows the protocol but may perform side calculations to breach security. While not the strongest assumption, this model fits real-world scenarios where trust is established before computation initiation.

\begin{figure}[t]
    \centering
      \includegraphics[width=1\linewidth]{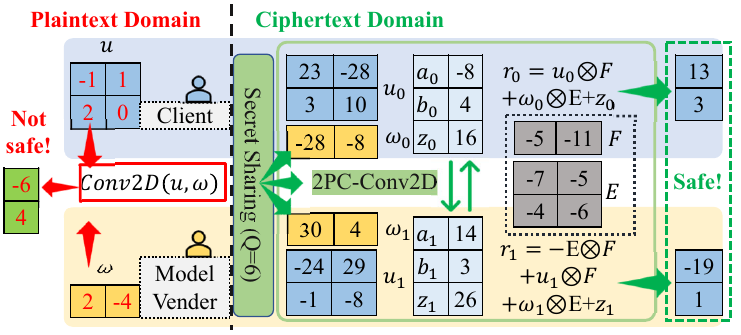}
  \captionof{figure}{Plaintext vs. ciphertext evaluation (4 bits). 
  }
    \label{fig:ss_example}
\end{figure}

\noindent\textbf{Secret Sharing Basics.}
As the most critical operation in multi-party computation, secret sharing bridges the communication between parties while still keeping one's information secure without the risk of being reasoned by other parties. Specifically, in this work, we adopt the commonly used secret sharing scheme described in CrypTen~\cite{knott2021crypten}. An example is given in Fig.~\ref{fig:ss_example}. As a symbolic representation, $\share{x}\gets(x_{S_0}, x_{S_1})$ denotes the two secret shares, where $x_{S_i}, i\in \{0,1\}$, is the share distributed to server $i$. The share generation and the share recovering adopted in our work are shown below: 
\begin{itemize}
\vspace{-3pt}
    \item {\it Share Generation} $\mathbb{\textrm{shr}} (x)$: A random value $r$ in $\mathbb{Z}_{m}$ is sampled, and shares are generated as $\share{x}\gets (r, x-r)$.
    \vspace{-3pt}
    \item {\it Share Recovering}  $\mathbb{\textrm{rec}} ({\share{x}})$: Given $\share{x}\gets (x_{S_0}, x_{S_1})$, it computes $x\gets x_{S_0} + x_{S_1}$ to recover $x$.
\end{itemize}
As most operators used in DNNs can be implemented through scaling, addition, multiplication, and comparison, here we provide an overview of these basic operations. 

\begin{figure*}[ht]
    \centering
      \includegraphics[width=.98\linewidth]{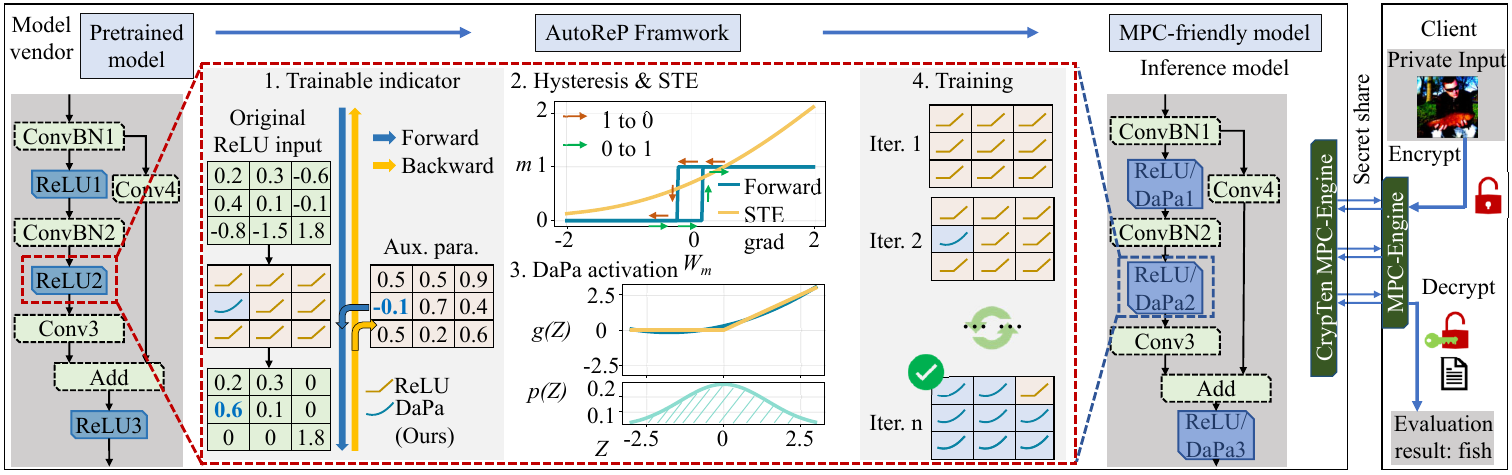}
  \captionof{figure}{Overview of \ourframework framework for 2PC DNN based private inference setup. 
  }
    \label{fig:framework}
\end{figure*}

\noindent\textbf{Scaling and Addition.} We denote secret shared matrices as $\share{X}$ and $\share{Y}$. The encrypted evaluation is given in Eq.~\ref{eq:mat_ad_ss}, where $a$ is the scaling factor. 
\begin{equation}\label{eq:mat_ad_ss}
\share{aX+Y}\gets(aX_{S_0}+Y_{S_0}, aX_{S_1}+Y_{S_1})
\vspace{-6pt}
\end{equation}
\noindent\textbf{Multiplication.}
In our work, we consider the use of matrix multiplicative operations in the secret-sharing pattern, specifically $\share{R}\gets \share{X} \otimes \share{Y}$, where $\otimes$ is a general multiplication such as Hadamard product, matrix multiplication, and convolution. To generate the required Beaver triples~\cite{beaver1991efficient} $\share{Z}=\share{A}\otimes\share{B}$, we utilize an oblivious transfer (OT)~\cite{kilian1988founding} based approach, with $A$ and $B$ being randomly initialized. It is important to ensure that the shapes of $\share{Z}, \share{A}$, and $\share{B}$ match those of $\share{R}, \share{X}$, and $\share{Y}$, respectively, in order to align the matrix computation. Next, each party computes two intermediate matrices, $E_{S_i} = X_{S_i} - A_{S_i}$ and $F_{S_i} = Y_{S_i} - B_{S_i}$, separately. The intermediate shares are then jointly recovered, with $E\gets \mathbb{\textrm{rec}} {(\share{E})}$ and $F\gets \mathbb{\textrm{rec}} {(\share{F})}$. Finally, each server $S_i$ calculates the secret-shared $R_{S_i}$ locally to get the result:
\begin{equation}\label{eq:mat_mul_ss}
R_{S_i} = -i\cdot E \otimes F + X_{S_i} \otimes F  + E \otimes Y_{S_i} + Z_{S_i}
\end{equation}

\noindent\textbf{Secure 2PC Comparison.}
In the context of secure MPC, the 2PC comparison protocol, also known as the millionaires' protocol, is designed to determine which of two parties holds a larger value, without revealing the actual value to each other. We uses the same protocol as CrypTen~\cite{knott2021crypten} to conduct comparison ($\share{X < 0}$) through \textcircled{1} arithmetic share $\share{X}$ to binary share $\bshare{X}$ conversion, \textcircled{2} right shift to extract the sign bit $\bshare{b} = \bshare{X} >> (L-1)$ ($L$ is the bit width), and \textcircled{3} binary share $\bshare{b}$ to arithmetic share $\share{b}$ conversion for final evaluation result.

\noindent\textbf{Ciphertext Square Operator}. Considering the element-wise square operator presented in Eq.~\ref{eq:sq_op}, where $\otimes$ denotes the Hadamard product, it is necessary to generate a Beaver pair, $\share{Z}$ and $\share{A}$, such that $\share{Z} = \share{A} \otimes \share{A}$. The pair $\share{A}$ is randomly initialized and shared among distinct parties through the use of oblivious transfer.

\begin{equation}\label{eq:sq_op}
\share{R}\gets \share{X} \otimes \share{X}
\end{equation}


Subsequently, the parties compute $\share{E} = \share{X} - \share{A}$ and collaboratively reconstruct $E$ using the recovery operation, $E \gets \mathbb{\textrm{rec}}(\share{E})$. The outcome, $R$, can be derived through the application of Eq.~\ref{eq:sq_eval}.

\begin{equation}\label{eq:sq_eval}
R_{S_i} = Z_{S_i} + 2 E \otimes A_{S_i}  + E \otimes E
\end{equation}

\subsection{Prior Arts Towards PI Acceleration.}
TAPAS~\cite{sanyal2018tapas} and XONN~\cite{riazi2019xonn} compressed the model into binary neural network format which has reduced number of bits, and the method can effectively reduce the size of Garbled circuit for MPC comparison protocol implementation. CryptoNAS~\cite{ghodsi2020cryptonas}, Sphynx~\cite{cho2021sphynx} and SafeNet~\cite{lou2020safenet} are typical NAS directed works which define searched spaces with reduced count of ReLU find the suitable architecture. Delphi~\cite{mishra2020delphi} focuses on partially replacing ReLU function with low order polynomial function to achieve speedup in MPC based PI. SNL~\cite{cho2022selective} and DeepReduce~\cite{jha2021deepreduce} developed frameworks to replace ReLUs with linear function. 
DeepReduce~\cite{jha2021deepreduce} involves manual design of neural architecture while SNL~\cite{cho2022selective} automates ReLU reduction process. Both prior works follow setting similar to Delphi~\cite{mishra2020delphi}, and exhibit higher 2PC comparison overhead than CrypTen~\cite{knott2021crypten} framework which is adopted in our research. 

%% file: Sections/Sec3_AutoReP.tex
\section{The \ourframework Framework}\label{sec:autorep_framework}

As depicted in Fig.~\ref{fig:framework}, we present \ourframework, an automatic replacement approach for accelerating DNN on PI while minimizing the inference accuracy drop. Our approach addresses the challenge of replacing the communication expensive non-linear activation function (i.e., ReLU) with PI-friendly low-order polynomial functions from pre-trained DNNs with less accuracy drop. 


We formulate ReLU replacement as a fine-grained feature-level optimization problem. Our solution involves a discrete indicator parameter that determines which ReLU operations should be replaced by polynomial functions to achieve minimal accuracy drop, which will be updated according to a hysteresis function~\cite{meng2020training}. 
Our approach improves upon SNL~\cite{cho2022selective} by training a discrete indicator parameter until both indicator and model weight converge, leading to superior convergence accuracy, as opposed to training a continuous slope parameter and fine-tuning the model after fixing the ReLU-polynomial selection in SNL~\cite{cho2022selective}. 


We propose an approach (DaPa) to determine the suitable polynomial activation functions to replace ReLU, based on the channel-wise feature map distribution. The combination of these techniques enables automatic and efficient acceleration of DNNs for PI for more than 10 times speedup. 


\subsection{Problem Formulation}
Our approach is generalizable to the replacement of ReLU activation functions in any $L$-layer differentiable neural networks $f_W$ parameterized by $W:= \{W_i\}_{i=0}^{L-1}$,
where the input $X_0 \in R^{m\times n}$ is mapped to the target $Y \in R^{d}$. 
Our goal is to replace the ReLU (denoted as $g_{r}$) with the polynomial function (denoted as $g_{p}$), with the aim of achieving an overall $N$ remaining ReLUs with minimal accuracy drop. 

We utilize an indicator parameter $m$ to indicate the replacement position on feature map-level: $m_{i \: k} = 0$ ($k_{th}$ element of $i_{th}$ layer), $g_{r}$ is replaced by $g_{p}$; $m_{i \: k} = 1$, use $g_{r}$. The proposed element-wise discrete indicator parameter $m_{i \: k}$ gives the expression of the $i_{th}$ layer with partially replaced ReLU as $X_{i \: k} = m_{i \: k} \odot g_{r}(Z_{(i-1) \: k}) + (1-m_{i \: k}) \odot g_{p}(Z_{(i-1) \: k})$, where $Z_{i-1}$ is the $i_{th}$ layer output. 
The problem of ReLU replacement can be formulated as follows:

{\small
\begin{equation}\label{eq:rep_problem}
\begin{split}
\underset{W}{argmin}  (\mathcal{L}(f_W(X_0), Y) 
+ \mu \cdot \max (\sum_{i = 1}^{L-1}||m_i||_0 - N), 0 ))
\end{split}
\end{equation}
}

 Where $\mathcal{L}$ denotes the loss function.
However, the second term of Eq.~\ref{eq:rep_problem} is non-differentiable due to the zero norm regularization applied on discrete indicator parameter $m_i$, making the problem intractable using traditional gradient-based optimization methods. To circumvent the non-differentiable behavior of the discrete indicator parameter, we introduce the utilization of a trainable auxiliary parameter $m_{W}$ to parameterize the discrete indicator parameter, as represented by Eq.~\ref{eq:para_wm}. However, Eq.~\ref{eq:para_wm} is still a step function and is non-differentiable. To approximate the gradient of Eq.~\ref{eq:para_wm}, we adopt straight-through estimator (STE) method which will be discussed in Sec.~\ref{sec:indicator_update}. 

{\small
\begin{equation}\label{eq:para_wm}
m_{i\: k} = m_{W, i\: k} > 0
\end{equation}
}

To analyze Eq.~\ref{eq:rep_problem}, we decompose the gradient of auxiliary parameters from Eq.~\ref{eq:rep_problem} into two parts: accuracy gradient ( Eq.~\ref{eq:Back_update}) from accuracy loss $\mathcal{L}_{acc}$ and ReLU count regularization gradient (Eq.~\ref{eq:Back_update_l2}) from ReLU count penalty $\mathcal{L}_{N}$. Eq.~\ref{eq:Back_update_l2} penalizes the auxiliary parameters based on the difference between $g_r(Z)$ and $g_p(Z)$, the term provides recoverability as it allows both gradient directions. Eq.~\ref{eq:Back_update_l2} penalizes the ReLU count and ensures the target number of ReLUs is met. 

{\small
\begin{equation}\label{eq:Back_update}
\frac{\partial \mathcal{L}_{acc}}{\partial m_{W, i\: k}} = \frac{\partial \mathcal{L}_{acc}}{\partial X_{i\: k}} (g_{r}(Z_{i-1}) - g_{p}(Z_{i-1})) \frac{\partial m_{i\: k}}{\partial m_{W, i\: k}} 
\end{equation}
\begin{equation}\label{eq:Back_update_l2}
\frac{\partial \mathcal{L}_{N}}{\partial m_{W, i\: k}} =  \left\{ \begin{array}{cc}
\mu \frac{\partial m_{i\: k}}{\partial m_{W, i\: k}},  &\textnormal{$||m_i||_0 - N > 0$} \\
0,  &\textnormal{otherwise} \\
\end{array} \right.
\end{equation}
}




\subsection{Update Rule of Indicator Parameter}
\label{sec:indicator_update}

\begin{figure}[t]
    \centering
      \includegraphics[width=0.95\linewidth]{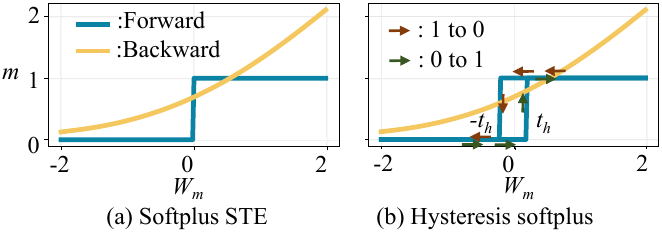}
  \caption{Hysteresis indicator parameter update. }
    \label{fig:hysteresis}
\end{figure}

\textbf{Indicator Parameter Gradient.} 
There are various STE functions for estimating the discrete function's gradient. While linear STE has been used in previous work~\cite{srinivas2017training}, recent studies suggest that ReLU-like STE has superior convergence \cite{yin2019understanding}. However, ReLU STE may cause gradient freezing problem~\cite{xiao2019autoprune}. To address this issue, we adopt the softplus function ($f(x) = log(1+e^x)$) based STE proposed in ~\cite{xiao2019autoprune}, which is shown in Fig.~\ref{fig:hysteresis}(a), to estimate the indicator parameter gradient $\frac{\partial m_{i: k}}{\partial W_{m, i: k}}$.

\textbf{Stability of Indicator Parameter Update.}
During each training iteration, the auxiliary parameter and indicator parameter updates are performed in accordance with Eq.~\ref{eq:axu_update} with softplus STE. However, there is a potential for indicator parameter instability issues to arise during forward binarization step as the training converges. This instability can occur when some of the $W_{m, i\: k}$ values are close to zero. Any small perturbation of these values during the update process may result in a flip of the indicator $m_{i\: k}$, thus affecting the training performance.

{\small
\begin{equation}\label{eq:axu_update}
 m_{W, i\: k} \mathrel{+}= \eta \frac{\partial \mathcal{L}_{2}}{\partial m_{W, i\: k}}, m_{ij} = m_{W, i\: k} > 0
\end{equation}
}

We propose the use of a hysteresis indicator parameter update to enhance the stability of the indicator parameter $m_{ij}$ during the forward binarization process with Eq.~\ref{eq:axu_update}. 

The proposed hysteresis indicator parameter update is depicted in Fig.~\ref{fig:hysteresis}(b). To reduce the possibility of the indicator flip, the hysteresis indicator parameter update utilizes the threshold $t_h$ as a hyperparameter and iteratively evaluates the old indicator values $m_{i\: k}$ and the updated auxiliary parameter $m_{W, i\: k}$ values to determine the new indicator values $m_{i\: k}$, as outlined in Eq.~\ref{eq:hysteresis_eq}. The hysteresis indicator parameter $m_{i\: k}$ reaches convergence when all auxiliary parameters $m_{W, i\: k}$ no longer fluctuate across the adjustable threshold $\pm t_h$.

{\small
\begin{equation}\label{eq:hysteresis_eq}
m_{i\: k, t+1} =  \left \{ \begin{array}{cccc}
1,  &\textnormal{$m_{i\: k, t} = 1$ and $m_{W, i\: k} > -t_h$} \\
0,  &\textnormal{$m_{i\: k, t} = 1$ and $m_{W, i\: k} \leq -t_h$} \\
0,  &\textnormal{$m_{i\: k, t} = 0$ and $m_{W, i\: k} \leq t_h$} \\
1,  &\textnormal{$m_{i\: k, t} = 0$ and $m_{W, i\: k} > t_h$}\\
\end{array} \right.
\end{equation}
}

\begin{figure}[ht]
    \centering
      \includegraphics[width=0.95\linewidth]{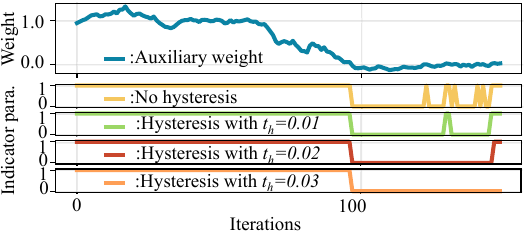}
  \caption{Balance the recoverability and stability through tuning hysteresis threshold}
    \label{fig:hysteresis_example}
\end{figure}

\begin{figure*}
    \centering
    \begin{minipage}[b]{0.245\textwidth}
        \centering
        \includegraphics[width=\linewidth]{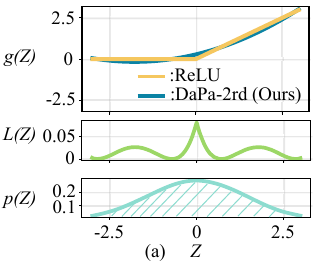}
    \end{minipage}
    \begin{minipage}[b]{0.21\textwidth}
        \centering
        \includegraphics[width=\linewidth]{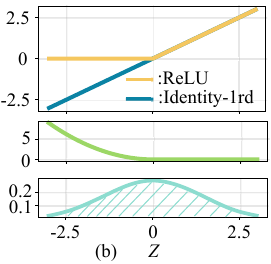}
    \end{minipage}
    \begin{minipage}[b]{0.21\textwidth}
        \centering
        \includegraphics[width=\linewidth]{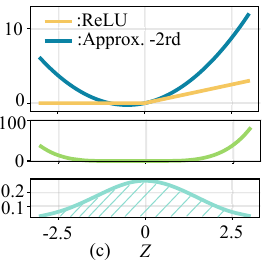}
    \end{minipage}
    \begin{minipage}[b]{0.21\textwidth}
        \centering
        \includegraphics[width=\linewidth]{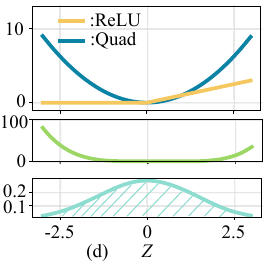}
    \end{minipage}
    \caption{ReLU replacement comparison with $N(0, 2)$ normal distribution. $L(Z)$ donates square error of approximation. (a) Proposed second order approximation: $g= 0.14Z^2 + 0.5Z + 0.28$. (b) Identity~\cite{cho2022selective}: $g= Z$. (c) Second order~\cite{ali2020polynomial}: $g= Z^2 + Z$. (d) Second order~\cite{mishra2020delphi, ali2020polynomial, garimella2021sisyphus}: $g= Z^2$. 
    }
    \label{fig:poly_norm2}
\end{figure*}

\textbf{Recoverability and Stability.}
The proposed update rule for the indicator parameter improves the recoverability and stability of the automatic ReLU replacement process. 
During the replacement, the ReLU will be replaced according to the difference between $g_r(Z)$ and $g_p(Z)$. The less difference between $g_r(Z)$ and $g_p(Z)$ on a feature map location, the more likely $g_r(Z)$ will be replaced by $g_p(Z)$.
However, important non-linear features $g_r(Z)$ may be replaced due to the ReLU count penalty.
Without recoverability, the replacement process will be more likely to be trapped in local optima (will be shown in experiment). The recoverability in this process is achieved through the use of the accuracy loss gradient in Eq.~\ref{eq:Back_update} combined with the softplus STE function. This automatic recovery process increases the accuracy of the ReLU replacement process without the need for an additional hard threshold, unlike in previous methods such as \cite{cho2022selective, guo2016dynamic}. However, unlike weight pruning, the non-linearity replacement process can also be unstable as training converges. To address this issue, we use the hysteresis indicator parameter update to improve the replacement stability.

An example is demonstrated in Fig.~\ref{fig:hysteresis_example}, the balance between recoverability and stability in the system can be controlled by tuning the hyperparameter $t_h$. A higher threshold results in decreased recoverability but increased stability, whereas a lower threshold results in the opposite effect. The optimal threshold selection, as shown in the example, lies within the range of 0.01 to 0.02. With the combination of learning rate decay and the hysteresis indicator parameter, the replacement process is able to balance exploration and exploitation throughout the training process, thus leading to a higher accuracy. 

\subsection{Polynomial Approximations of ReLU}\label{sec:dapa}



Low-order polynomial functions, such as first-order and second-order polynomials, can significantly reduce latency and communication volume in DNN PI applications by more than 20 times~\cite{mishra2020delphi}. 
However, first-order polynomial function (linear) does not provide non-linearity, so it could lead to significant model  expressivity reduction and lower accuracy under a high reduction ratio when used as ReLU replacement. In contrast, the second-order polynomial provides a certain degree of non-linearity and might be a better replacement for ReLU. 
However, prior works\cite{mishra2020delphi, lou2020safenet, ali2020polynomial, garimella2021sisyphus} have yet to find an effective method for determining the coefficient of the second-order polynomial functions for ReLU replacement. 
To improve the performance and training stability of ReLU replacement, we propose a feature map distribution-aware polynomial approximation (DaPa) for our \ourframework framework.
\textbf{Distribution-aware approximation.} 
The intuition behind the distribution-aware approximation is that minimizing the discrepancy between the output before and after ReLU replacement would lead to a smaller decrease in accuracy. This can be achieved by minimizing the minimum square error (MSE) between $g_{r}(Z)$ and $g_{p}(Z)$:

{\small
\begin{equation}\label{eq:poly_loss_org}
 \underset{w}{min}L(g_{p}) = \underset{g_{p}}{min}\sum (g_{r}(Z) - g_{p}(Z))^2
\end{equation}
}

Previous approaches~\cite{cho2022selective, ali2020polynomial, mishra2020delphi, garimella2021sisyphus} adopt a fixed polynomial function for all layers without taking feature map distribution into account, and leads to worse accuracy and training stability. In contrast to those approaches, we propose a distribution-aware approximation method to dynamically adjust the parameters of the polynomial function. Specifically, for a polynomial function of degree $s$ ($s > 0$), denoted as $g_{p, s}(Z, c) = \sum_{i = 0}^{s}c_i Z^i$, where $Z$ and $c_i$ are input and polynomial coefficient, the optimization problem in Eq.~\ref{eq:poly_loss_org} can be solve by getting the input probability distribution $p(Z)$, and reformulated the problem as follows: 

{\small
\begin{equation}\label{eq:poly_loss}
 \underset{c}{min}L(c) = \underset{c}{min}\int (g_{r}(Z) - g_{p, s}(Z, c))^2 p(Z) dZ
\end{equation}
}

Eq.~\ref{eq:poly_loss} can be solved numerically through the Monte Carlo integration and sampling. 
Here we present an analytical expression of the approximation of the ReLU activation function using second-order polynomial functions. The input $Z$ is assumed to be drawn from a normal distribution with mean $\mu$ and variance $\sigma^2$, i.e., $Z\sim N(\mu, \sigma^2)$. The analytical solution that minimizes Eq.~\ref{eq:poly_loss} is shown in Eq.~\ref{eq:poly_2rd_sol}. 

{\small
\begin{equation}\label{eq:poly_2rd_sol}
\begin{split}
 c_0 = & \frac{\sqrt{2}\mu^2e^{-\frac{\mu^2}{2\sigma^2}}}{4\sqrt{\pi}\sigma} + \frac{\sqrt{2}\sigma e^{-\frac{\mu^2}{2\sigma^2}}}{4\sqrt{\pi}},
c_2 = \frac{\sqrt{2}e^{-\frac{\mu^2}{2\sigma^2}}}{4\sqrt{\pi}\sigma},\\ 
&c_1 = -\frac{\sqrt{2}\mu e^{-\frac{\mu^2}{2\sigma^2}}}{2\sqrt{\pi}\sigma} - \frac{erfc(\frac{\sqrt{2}\mu}{2\sigma})}{2} + 1
\end{split}
\end{equation}
}

The minimum approximation loss is given in Eq.~\ref{eq:poly_2rd_min_loss}. 
{\small
\begin{equation}\label{eq:poly_2rd_min_loss}
\begin{split}
\underset{c}{min}L & = -\frac{\mu^2 \operatorname{erfc}^2\left(\frac{\sqrt{2} \mu}{2 \sigma}\right)}{4}+ \frac{\mu^2 \operatorname{erfc}\left(\frac{\sqrt{2} \mu}{2 \sigma}\right)}{2}+ \\ 
& \frac{\sqrt{2} \mu \sigma e^{-\frac{\mu^2}{2 \sigma^2}} \operatorname{erfc}\left(\frac{\sqrt{2} \mu}{2 \sigma}\right)}{2 \sqrt{\pi}}-\frac{\sqrt{2} \mu \sigma e^{-\frac{\mu^2}{2 \sigma^2}}}{2 \sqrt{\pi}}- \\ 
& \frac{\sigma^2 \operatorname{erfc}^2\left(\frac{\sqrt{2} \mu}{2 \sigma}\right)}{4}+\frac{\sigma^2 \operatorname{erfc}\left(\frac{\sqrt{2} \mu}{2 \sigma}\right)}{2}-\frac{3 \sigma^2 e^{-\frac{\sigma^2}{\sigma^2}}}{4 \pi}
\end{split}
\end{equation}
}

\begin{table*}[t]
\caption{Baselines accuracy, ReLU counts, and latency}
\centering
\resizebox{0.9\linewidth}{!}{
\begin{tabular}{c|c|c|c|c|c|c|c}
\hline
Datasets      & \multicolumn{2}{c|}{CIFAR-10} & \multicolumn{2}{c|}{CIFAR-100} & \multicolumn{2}{c|}{Tiny-ImageNet} & ImageNet \\ \hline
Models        & ResNet-18 & WideResNet-22-8 & ResNet-18 & WideResNet-22-8 & ResNet-18 & WideResNet-22-8 & EfficientNet-B2 \\ \hline
Accuracy (\%) & 95.55 & 96.29 &77.8 & 80.2  &65.48 & 66.77 & 78.158 \\ \hline
ReLUs (K) &  491.52 & 1359.87 & 491.52 & 1359.87  & 1966.08 & 5439.488 & 8804.162  \\ \hline
Latency (s) & 1.234 & 3.219  & 1.242  & 3.230  & 3.219  & 7.978  & 11.276 \\ \hline
\end{tabular}
}
\label{tab:baseline}

\end{table*}

We illustrate the effectiveness of our proposed approximation method with a second-order approximation of ReLU function for inputs drawn from a normal distribution $Z\sim N(0, 2)$. It is important to note that the feature map distribution may vary, and this is just an illustrative example.
These results are illustrated in Fig.~\ref{fig:poly_norm2}. 
It can be observed that the proposed second-order approximation results in the lowest error to approximate ReLU function. 

\textbf{Channel-wise approximation.} 
For most CNNs, the distribution of intermediate feature map follows a channel-wise manner due to the batch normalization module. Inspired by this fact, we propose a more fine-grained and accurate ReLU replacement method by adopting channel-wise polynomial approximation. Unlikely prior works which approximate the ReLU function using identical polynomial function across entire CNNs~\cite{cho2022selective, mishra2020delphi, ali2020polynomial, ali2020polynomial, garimella2021sisyphus}, the proposed channel-wise approximation gives a smaller accuracy loss. 

%% file: Sections/Sec4_experiment.tex
\section{Experiments}

\subsection{Experimental Setup}


\noindent\textbf{PI system setup.} 
Our platform comprises two servers equipped with RTX6000, which are connected to a router with a bandwidth of 1 GB/s via a local area network (LAN). To implement secure computation for PI, we utilize the CrypTen~\cite{knott2021crypten} framework with the admissible adversary assumption~\cite{mohassel2017secureml} (refer to Sec.~\ref{sec:background}).

\noindent\textbf{Architectures and datasets} 
To enable cross-work comparison with state-of-the-art approaches, we evaluate \ourframework using second-order polynomial replacement on ResNet-18~\cite{he2016deep} and WideResNet-22-8~\cite{zagoruyko2016wide} architectures on CIFAR-10/CIFAR-100~\cite{krizhevsky2009learning} and Tiny-ImageNet~\cite{chrabaszcz2017downsampled} datasets. To ensure a fair comparison with previous works~\cite{jha2021deepreduce, cho2022selective}, we remove ReLU layers from the first convolutional layer. For scalability evaluation, we apply \ourframework to EfficientNet-B2~\cite{tan2019efficientnet} with ReLU activation function on ImageNet~\cite{krizhevsky2012imagenet}. See Table~\ref{tab:datasets} for more dataset information.

\begin{table}[ht]
\caption{Image classification datasets}
\centering
\resizebox{0.98\linewidth}{!}{
\begin{tabular}{ccccc}
\hline
Dataset & Image size & Class & \begin{tabular}[c]{@{}c@{}}Training Samples\\ per class\end{tabular} & \begin{tabular}[c]{@{}c@{}}Test Samples\\ per class\end{tabular} \\ \hline
CIFAR-10~\cite{krizhevsky2009learning}      & 32 × 32    & 10    & 5000                                                                 & 1000                                                             \\ \hline
CIFAR-100~\cite{krizhevsky2009learning}     & 32 × 32    & 100   & 500                                                                  & 100                                                              \\ \hline
Tiny-ImageNet~\cite{chrabaszcz2017downsampled} & 64 × 64    & 200   & 500                                                                  & 50                                                               \\ \hline
ImageNet~\cite{krizhevsky2012imagenet}      & 224 × 224  & 1000  & $\sim$1282                                                           & 50                                                               \\ \hline
\end{tabular}
}
\label{tab:datasets}
\end{table}

\noindent\textbf{Baselines.} 
For ResNet-18 and WideResNet-22-8 on CIFAR-10/100 and Tiny-ImageNet datasets, we pre-train the models using SGD with an initial learning rate (LR) of 0.1 and momentum of 0.9 for 400 epochs. The LR is scheduled using a standard cosine annealing LR scheduler. In the case of EfficientNet-B2 trained on ImageNet, we utilize the PyTorch pre-trained weights~\cite{torchvision}. As the EfficientNet-B2 model uses SiLU as the default non-linear activation function, we replace the SiLU activation with ReLU and fine-tuned the model using SGD with a LR of 0.01, momentum of 0.9, and cosine annealing LR scheduler. 
See Table~\ref{tab:baseline} for the accuracy and number of ReLUs for the baseline models.


\noindent\textbf{\ourframework algorithm settings} 
We employ the \ourframework algorithm with a second-order polynomial replacement on the aforementioned pre-trained models. For the indicator parameter update, we use the Adam optimizer with a LR of 0.001, while for the model weight, we use the Adam optimizer with a LR of 0.0001. The LR for both parameters are scheduled to decay based on the cosine annealing decay function. We conduct the majority of the replacement experiments using 150 replacement epochs. To balance recoverability and stability, we set the hysteresis threshold to $t_h = 0.003$. The details of hysteresis threshold selection are in the ablation study (Section~\ref{sec:ablation}). As described in Section~\ref{sec:dapa}, we capture the channel-wise running mean and variance and determine the polynomial function's parameter based on Eq.~\ref{eq:poly_2rd_sol}.





\begin{table}[t]
\caption{Cross-work comparison on CIFAR-100}
\centering
\resizebox{0.98\linewidth}{!}{
\begin{tabular}{ccccc}
\hline
\multicolumn{1}{c|}{Dataset}        & \multicolumn{4}{c}{CIFAR-100}                                                                                                                                                       \\ \hline
\multicolumn{1}{c|}{Methods}        & \begin{tabular}[c]{@{}c@{}}\#ReLUs\\ (K)\end{tabular} & \begin{tabular}[c]{@{}c@{}}Test Acc.\\ (\%)\end{tabular} & \begin{tabular}[c]{@{}c@{}}Latency\\ (s)\end{tabular} & Acc./ReLU \\ \hline
\multicolumn{5}{c}{ReLU $\le$ 100 K}                                                                                                                                                                                         \\ \hline
\multicolumn{1}{c|}{CryptoNAS}      & 100.0                                                 & 68.5                                                    & 2.3                                                   & 0.685     \\
\multicolumn{1}{c|}{Sphynx}         & 51.2                                                  & 69.57                                                   & 1.335                                                 & 1.359     \\
\multicolumn{1}{c|}{Sphynx}         & 25.6                                                  & 66.13                                                   & 0.727                                                 & 2.583     \\
\multicolumn{1}{c|}{DeepReduce}     & 49.2                                                  & 69.5                                                    & 1.19                                                  & 1.413     \\
\multicolumn{1}{c|}{DeepReduce}     & 12.3                                                  & 64.97                                                   & 0.45                                                  & 5.283     \\
\multicolumn{1}{c|}{SNL\textsuperscript{+}}            & 49.9                                                  & 73.75                                                   & 1.066                                                 & 1.478     \\
\multicolumn{1}{c|}{SNL\textsuperscript{+}}            & 12.9                                                  & 66.53                                                   & 0.291                                                 & 5.517     \\
\multicolumn{1}{c|}{\textbf{AutoReP (Ours)\textsuperscript{+}}} & 50                                                    & 75.48                                                   & 0.252                                                 & 1.510     \\
\multicolumn{1}{c|}{\textbf{AutoReP (Ours)\textsuperscript{+}}} & 12.9                                                  & 74.92                                                   & 0.170                                                 & 5.808     \\
\multicolumn{1}{c|}{\textbf{AutoReP (Ours)\textsuperscript{+}}} & 6                                                     & 73.79                                                   & 0.155                                                 & 12.298    \\ \hline
\multicolumn{5}{c}{ReLU \textgreater 100 K}                                                                                                                                                                              \\ \hline
\multicolumn{1}{c|}{CryptoNAS}      & 344.0                                                 & 76.0                                                    & 7.50                                                  & 0.221     \\
\multicolumn{1}{c|}{Sphynx}         & 230.0                                                 & 74.93                                                   & 5.12                                                  & 0.326     \\
\multicolumn{1}{c|}{DeepReduce}     & 229.4                                                 & 76.22                                                   & 4.61                                                  & 0.332     \\
\multicolumn{1}{c|}{DeepReduce}     & 197.0                                                 & 75.51                                                   & 3.94                                                  & 0.383     \\
\multicolumn{1}{c|}{SNL\textsuperscript{*}}            & 180.0                                                 & 77.65                                                   & 4.054                                                 & 0.431     \\
\multicolumn{1}{c|}{SNL\textsuperscript{*}}            & 120.0                                                 & 76.35                                                   & 2.802                                                 & 0.636     \\
\multicolumn{1}{c|}{\textbf{AutoReP (Ours)\textsuperscript{*}}} & 180                                                   & 78.23                                                   & 0.679                                                 & 0.435     \\
\multicolumn{1}{c|}{\textbf{AutoReP (Ours)\textsuperscript{*}}} & 150                                                   & 78.38                                                   & 0.614                                                 & 0.523     \\
\multicolumn{1}{c|}{\textbf{AutoReP (Ours)\textsuperscript{*}}} & 120                                                   & 77.56                                                   & 0.550                                                 & 0.646     \\ \hline
\end{tabular}
}
\captionsetup{singlelinecheck=false, justification=raggedright, font=footnotesize, skip=5pt}
\caption*{\textsuperscript{+}: starts with ResNet-18. \:   
\textsuperscript{*}: starts with WideResNet-22-8. }
\label{tab:cifar100}
\end{table}

\begin{figure*}[ht]
    \centering
      \includegraphics[width=.95\linewidth]{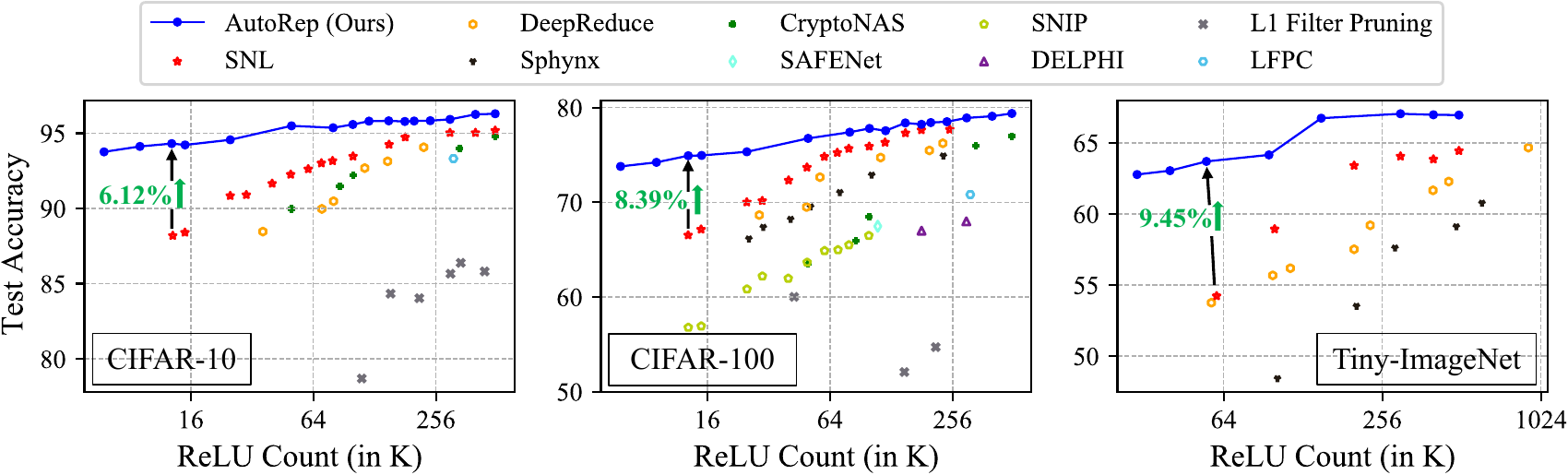}
  \captionof{figure}{AutoRep achieves Pareto frontiers of ReLU counts vs. test accuracy on CIFAR-10, CIFAR-100, and Tiny-ImageNet. AutoRep outperforms the state-of-the-art methods (SNL~\cite{cho2022selective}, DeepReDuce~\cite{jha2021deepreduce}, Sphynx~\cite{cho2022sphynx}, CryptoNAS~\cite{ghodsi2020cryptonas}, SAFENet~\cite{lou2020safenet}, SNIP~\cite{lee2018snip}, DELPHI~\cite{mishra2020delphi}, L1 filter pruning~\cite{li2016pruning} and LFPC~\cite{he2020learning}) in all range of ReLU counts on all datasets. 
  }
    \label{fig:Pareto_frontiers}
\end{figure*}

\subsection{Experimental Results}
\ourframework
is evaluated and compared with SOTA works ~\cite{cho2022selective, jha2021deepreduce, cho2022sphynx, ghodsi2020cryptonas, lou2020safenet, lee2018snip, mishra2020delphi, li2016pruning, he2020learning}, and results are presented in Fig.~\ref{fig:Pareto_frontiers}. Our proposed framework achieves significantly better results than the other approaches on the CIFAR-10, CIFAR-100, and Tiny-ImageNet datasets.

\subsubsection{Pareto Frontier and Cross-work comparison}
Table~\ref{tab:cifar100} and Table~\ref{tab:tinyimagenet} provide detailed information on the accuracy and latency trade-offs for different ReLU budgets on the CIFAR-100 and Tiny-ImageNet, respectively. It is worth noting that previous works~\cite{ghodsi2020cryptonas, cho2022sphynx, jha2021deepreduce, cho2022selective} use DELPHI~\cite{mishra2020delphi} as the PI framework, which employs a garbled circuit implementation that is more computationally and communicationally expensive than the CrypTen~\cite{knott2021crypten} used in our evaluation. As a result, their reported latencies may be higher than those presented in our study.

On the CIFAR-100 dataset using ResNet-18,  \ourframework achieves 74.92\% accuracy with 12.9K ReLU budget, which is 8.39\% higher than SNL~\cite{cho2022selective}, with the same ReLU budget, while reducing the inference latency by 1.7 times. 
Our experiments demonstrate that \ourframework has less accuracy drop for ReLU replacement compared to SNL~\cite{cho2022selective}. 
To achieve a similar accuracy level of 73.79\%, \ourframework requires only 6K ReLUs, resulting in an 8.2 times reduction in ReLU budget when compared to SNL~\cite{cho2022selective}.
For higher ReLU budgets, \ourframework on WideResNet-22-8 achieves an accuracy improvement of 0.7\% to 1.2\% compared to SNL~\cite{cho2022selective} on 120K to 180K ReLU counts, while also outperforming DeepReduce~\cite{jha2021deepreduce} with fewer ReLU budgets.


Our \ourframework achieves stronger performance than prior SOTA methods on the Tiny-ImageNet dataset.
Starting from ResNet-18 with a ReLU budget of 55K, our \ourframework achieves 63.69\% accuracy, outperforming SNL~\cite{cho2022selective} with a 59.1K ReLU budget by 9.45\%, while reducing the ReLU budget by a factor of 3.6 compared to SNL with a similar accuracy of 63.39\% on a ReLU budget of 198.1K. 
Moreover,  \ourframework achieves a 3.39 $\times$ latency reduction with a similar 55K ReLU budget, demonstrating that second-order polynomial replacement does not result in higher latency. 
For higher ReLU budgets, \ourframework starting from WideResNet-22-8 achieves more than 2.5\% average accuracy improvement compared to SNL~\cite{cho2022selective} and DeepReduce~\cite{jha2021deepreduce}.

\begin{table}[ht]
\caption{Cross-work comparison on Tiny-ImageNet}
\centering
\resizebox{0.98\linewidth}{!}{
\begin{tabular}{ccccc}
\hline
\multicolumn{1}{c|}{Dataset}        & \multicolumn{4}{c}{Tiny-ImageNet}                                                                                                                                                   \\ \hline
\multicolumn{1}{c|}{Methods}        & \begin{tabular}[c]{@{}c@{}}\#ReLUs\\ (K)\end{tabular} & \begin{tabular}[c]{@{}c@{}}Test Acc.\\ (\%)\end{tabular} & \begin{tabular}[c]{@{}c@{}}Latency\\ (s)\end{tabular} & Acc./ReLU \\ \hline
\multicolumn{5}{c}{ReLU $\le$ 100 K}                                                                                                                                                                                         \\ \hline
\multicolumn{1}{c|}{Sphynx}         & 102.4                                                 & 48.44                                                   & 2.35                                                  & 0.473     \\
\multicolumn{1}{c|}{DeepReduce}     & 57.35                                                 & 53.75                                                   & 1.85                                                  & 0.937     \\
\multicolumn{1}{c|}{SNL\textsuperscript{+}}            & 99.6                                                  & 58.94                                                   & 2.117                                                 & 0.592     \\
\multicolumn{1}{c|}{SNL\textsuperscript{+}}            & 59.1                                                  & 54.24                                                   & 1.265                                                 & 0.918     \\
\multicolumn{1}{c|}{\textbf{AutoReP (Ours)\textsuperscript{+}}} & 55                                                    & 63.69                                                   & 0.373                                                 & 1.158     \\
\multicolumn{1}{c|}{\textbf{AutoReP (Ours)\textsuperscript{+}}} & 30                                                    & 62.77                                                   & 0.335                                                 & 2.092     \\ \hline
\multicolumn{5}{c}{100 K \textless ReLU $\le$ 300 K}                                                                                                                                                                         \\ \hline
\multicolumn{1}{c|}{Sphynx}         & 204.8                                                 & 53.51                                                   & 4.401                                                 & 0.261     \\
\multicolumn{1}{c|}{DeepReduce}     & 196.6                                                 & 57.51                                                   & 4.61                                                  & 0.293     \\
\multicolumn{1}{c|}{SNL\textsuperscript{+}}            & 393.2                                                 & 61.65                                                   & 7.77                                                  & 0.157     \\
\multicolumn{1}{c|}{SNL\textsuperscript{+}}            & 204.8                                                 & 53.51                                                   & 4.401                                                 & 0.261     \\
\multicolumn{1}{c|}{\textbf{AutoReP (Ours)\textsuperscript{+}}} & 290                                                   & 64.74                                                   & 0.723                                                 & 0.223     \\
\multicolumn{1}{c|}{\textbf{AutoReP (Ours)\textsuperscript{+}}} & 190                                                   & 64.32                                                   & 0.574                                                 & 0.338     \\ \hline
\multicolumn{5}{c}{300 K \textless ReLU $\le$ 1000 K}                                                                                                                                                                        \\ \hline
\multicolumn{1}{c|}{Sphynx}         & 614.4                                                 & 60.76                                                   & 12.548                                                & 0.099     \\
\multicolumn{1}{c|}{DeepReduce}     & 917.5                                                 & 64.66                                                   & 17.16                                                 & 0.070     \\
\multicolumn{1}{c|}{SNL\textsuperscript{*}}            & 488.8                                                 & 64.42                                                   & 10.281                                                & 0.132     \\
\multicolumn{1}{c|}{\textbf{AutoReP (Ours)\textsuperscript{*}}} & 300                                                   & 67.04                                                   & 1.094                                                 & 0.223     \\ \hline
\end{tabular}
}
\captionsetup{singlelinecheck=false, justification=raggedright, font=footnotesize, skip=5pt}
\caption*{\textsuperscript{+}: starts with ResNet-18. \:   
\textsuperscript{*}: starts with WideResNet-22-8. }
\label{tab:tinyimagenet}
\end{table}


\subsubsection{\ourframework with  Linear Replacement}



\begin{figure}[ht]
    \centering
      \includegraphics[width=.99\columnwidth]{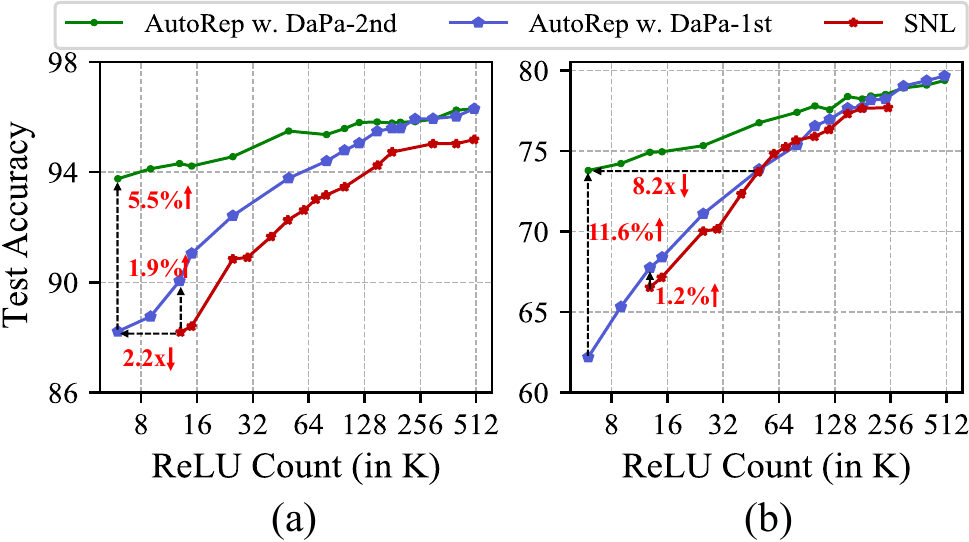}
  \captionof{figure}{\ourframework w. Dapa-2nd, \ourframework w. Dapa-1st, and SNL comparisons on (a) CIFAR-10 (b) CIFAR-100
  }
    \label{fig:Dapa_1st_vs_2nd}
\end{figure}

As discussed in Section~\ref{sec:autorep_framework}, \ourframework outperforms previous works in two key aspects: (1) fine-grained replacement policy using discrete indicator parameter and (2) DaPa activation function. To evaluate the contribution of each aspect, we conduct experiments using a first-order polynomial function and compare it with \ourframework using a second-order DaPa function and SNL~\cite{cho2022selective}. The evaluation results on CIFAR-10 and CIFAR-100 datasets are given in Fig.~\ref{fig:Dapa_1st_vs_2nd}. 
Note that \ourframework and  SNL use a similar DNN setting.



Our experiments on CIFAR-10 demonstrate that \ourframework with first-order polynomial replacement achieves a 90.05\% accuracy with 12.9K ReLU budgets, which is 1.86\% higher than SNL~\cite{cho2022selective}, but 4.3\% lower than \ourframework with second-order DaPa replacement. On average, \ourframework with first-order replacement yields 1.8\% higher accuracy than SNL~\cite{cho2022selective} under the same ReLU budgets across most of the range and achieves a 2.2 $\times$ reduction in ReLU budgets in extreme cases (6K vs. 12.9K).

For CIFAR-100, \ourframework with first-order polynomial replacement achieves 67.75\% accuracy under 12.9K ReLU budgets, which is 1.2\% higher than SNL~\cite{cho2022selective} under the same ReLU budgets, and 0.4\% higher than SNL~\cite{cho2022selective} with 15K ReLU budgets. However, the accuracy drop of \ourframework with first-order polynomial replacement is substantial compared to \ourframework with DaPa-2nd polynomial replacement, exhibiting an 11.6\% accuracy drop under the extreme case of low ReLU budgets (6K).

The experimental results indicate that our proposed \ourframework outperforms SNL~\cite{cho2022selective} in both first and second-order replacements, providing evidence to support the two main claims of our framework: \textcircled{1} better replacement policy, which leads to better convergence, and \textcircled{2} improved model expressivity through the deployment of second-order DaPa polynomial function.

\subsubsection{ImageNet Evaluation}
To showcase the effectiveness of \ourframework on larger models and datasets, we conduct experiments on EfficientNet-B2 with ImageNet using DaPa-2nd replacement. We run the replacement policy for 20 epochs with the same LR setting as previously. Results are shown in Table~\ref{tab:imagenet}. Compared to the baseline model's 78.158\% accuracy, \ourframework achieves a 17.6 $\times$ reduction in ReLU budget with only 1.592\% accuracy drop for the 500K ReLU case. Moreover, we achieve a 5.8 $\times$ speedup for private inference in this case. In the case of an extremely low ReLU budget of 50K, our \ourframework achieves a 176.1 $\times$ reduction in ReLU budget and a 7.8\% $\times$ speedup compared to the baseline model, with only a 2.61\% accuracy drop. The results demonstrate that our \ourframework with DaPa-2nd achieves a significant reduction in ReLU budget and inference speedup while preserving good model accuracy, even for relatively larger models and datasets.
\begin{table}[ht]
\caption{\ourframework for ImageNet}
\centering
\resizebox{0.91\linewidth}{!}{
\begin{tabular}{c|cccc}
\hline
Dataset        & \multicolumn{4}{c}{ImageNet}                                                                                                                                                         \\ \hline
Methods        & \begin{tabular}[c]{@{}c@{}}\#ReLUs\\ (K)\end{tabular} & \begin{tabular}[c]{@{}c@{}}Test Acc.\\ (\%)\end{tabular} & \begin{tabular}[c]{@{}c@{}}Latency \\ (s)\end{tabular} & Acc./ReLU \\ \hline
\textbf{AutoReP} & 500                                                   & 76.566                                                  & 1.945                                                  & 0.153     \\
\textbf{AutoReP} & 400                                                   & 76.368                                                  & 1.832                                                  & 0.191     \\
\textbf{AutoReP} & 300                                                   & 76.216                                                  & 1.72                                                   & 0.254     \\
\textbf{AutoReP} & 200                                                   & 76.176                                                  & 1.608                                                  & 0.381     \\
\textbf{AutoReP} & 100                                                   & 75.766                                                  & 1.495                                                  & 0.758     \\
\textbf{AutoReP} & 50                                                    & 75.548                                                  & 1.439                                                  & 1.511     \\ \hline
\end{tabular}
}
\captionsetup{singlelinecheck=false, justification=raggedright, font=footnotesize, skip=5pt}
\caption*{\: \: \: All start from EfficientNet-B2}
\label{tab:imagenet}
\end{table}



\subsection{Ablation Study}\label{sec:ablation}


\textbf{Threshold sensitivity}.
To investigate the impact of hyperparameters on the proposed \ourframework, we conduct experiments on ResNet-18 architecture trained on CIFAR-100 dataset while varying the hysteresis threshold $t_h$. Specifically, we compare the accuracy performance under different $t_h$ settings while keeping other hyperparameters fixed. 
\begin{figure}[ht]
    \centering
      \includegraphics[width=.95\columnwidth]{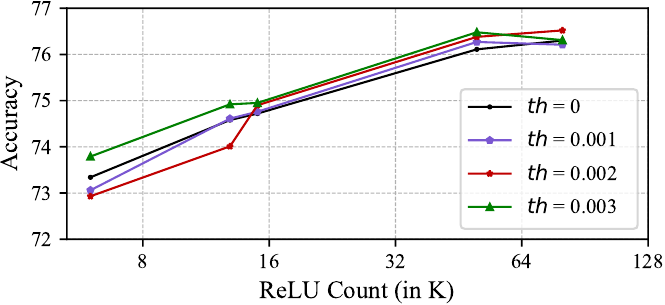}
  \captionof{figure}{\ourframework for ResNet-18 on CIFAR-100 dataset under different hysteresis thresholds setting. }
    \label{fig:threshold}
\end{figure}
The results are presented in Fig.~\ref{fig:threshold}. A lower hysteresis threshold may cause the binarized indicator parameter to flip frequently during convergence, leading to a decrease in accuracy. A hysteresis threshold of 0 cause  a decrease in accuracy of 0.45\% compared to a threshold of 0.003 under 6K ReLU budgets. Our experiments show that $t_h = 0.003$ strikes a good balance between the stability of the binarized indicator parameter and recoverable training, resulting in higher overall accuracy under most cases. Therefore, we adopt $t_h = 0.003$ for other experiments.

\begin{figure}[!h]
    \centering
      \includegraphics[width=.95\columnwidth]{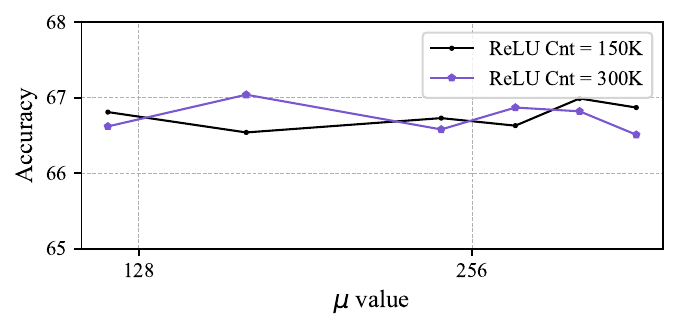}
  \captionof{figure}{\ourframework for Wide-ResNet-22-8 on Tiny-ImageNet dataset with different penalty $\mu$. }
    \label{fig:parameter_sensitivity}
\end{figure}

\textbf{Parameter sensitivity}.
To substantiate the assertion that the training of the proposed \ourframework is robust to variations in the ReLU count penalty parameter $\mu$, as stated in the primary problem formulation of the paper, we conduct a parameter sensitivity analysis. The results are depicted in Fig.~\ref{fig:parameter_sensitivity}, where $\mu$ is normalized by multiplying it with the original number of ReLUs. Our findings reveal that, under the same training setup, the overall accuracy exhibits only minor fluctuations across different values of $\mu$. This implies that careful tuning of the $\mu$ parameter is not necessary for the \ourframework framework to achieve good performance.



%% file: Sections/Sec5_conclusion.tex
\section{Conclusion}

Existing MLaaS (Machine Learning as a Service) accelerations have focused on plaintext inference acceleration~\cite{peng2022towards, wu2020intermittent, sheng2022larger, kan2021zero, luo2022codg, li2021generic, zhang2022toward, huang2021sparse, li2022makes, kan2022brain, peng2021accelerating, zhang2022algorithm, bao2019efficient, wang2023digital, huang2022automatic, wang2021lightweight, huang2021hmc, kan2022fbnetgen, xiao2019autoprune, peng2021binary, qi2021accelerating, qi2021accommodating, zhang2023accelerating, bao2020fast, peng2022length}. Others target plaintext training acceleration~\cite{wu2021enabling, wu2020enabling, wu2022decentralized, huang2023neurogenesis, bao2022accelerated, wu2023synthetic, bao2022doubly, xu2023neighborhood, wu2022distributed, lei2023balance, huang2022dynamic}, federated learning in order to protect the privacy of training data~\cite{wang2022variance, wu2021federated2, wu2021federated1}, and privacy protection for model vendors~\cite{wang2022analyzing, wang2020against}.

We propose the \ourframework framework, designed to be seamlessly integrated into MPC-based PI systems for MLaaS provider, and compatible with pre-trained CNN models on datasets of varying scales. The framework's fine-grained ReLU replacement policy and the distribution-aware polynomial approximation (DaPa) activation function enable it to achieve a 74.92\% accuracy on the CIFAR-100 dataset with 12.9K ReLU budget, outperforming the SOTA SNL~\cite{cho2022selective} framework by 8.39\%. \ourframework achieves 75.55\% accuracy when applied to EfficientNet-B2 on ImageNet and achieve a 176.1 $\times$ ReLU reduction. 

\section*{Acknowledgement}
This work was in part supported by the NSF CNS-2247891, 2247892, 2247893, CNS-2153690, CNS-2239672, US DOE Office of Science and Office of Advanced Scientific Computing Research under award 66150: "CENATE - Center for Advanced Architecture Evaluation". The Pacific Northwest National Laboratory is operated by Battelle for the U.S. Department of Energy under Contract DE-AC05-76RL01830. Any opinions, findings, conclusions, or recommendations expressed in this material are those of the authors and do not necessarily reflect the views of the funding agencies.

%